# A mathematical model of the metabolism of a cell. Self-organization and chaos

V.I. Grytsay[a,*], I.V. Musatenko[b]

[a]**Bogolyubov Institute for Theoretical Physics, Kyiv, Ukraine**
E-mail: **vgrytsay@bitp.kiev.ua**
[b] **Taras Shevchenko National University of Kyiv, Faculty of Cybernetics, Kyiv, Ukraine**
E-mail: ivmusatenko@gmail.com

**Abstract:** Using the classical tools of nonlinear dynamics, we study the process of self-organization and the appearance of the chaos in the metabolic process in a cell with the help of a mathematical model of the transformation of steroids by a cell *Arthrobacter globiformis*. We constructed the phase-parametric diagrams obtained under a variation of the dissipation of the kinetic membrane potential. The oscillatory modes obtained are classified as regular and strange attractors. We calculated the bifurcations, by which the self-organization and the chaos occur in the system, and the transitions "chaos-order", "order-chaos", "order-order," and "chaos-chaos" arise. Feigenbaum's scenarios and the intermittences are found. For some selected modes, the projections of the phase portraits of attractors, Poincaré sections, and Poincaré maps are constructed. The total spectra of Lyapunov indices for the modes under study are calculated. The structural stability of the attractors is demonstrated. A general scenario of the formation of regular and strange attractors in the given metabolic process in a cell is found. The physical nature of their appearance in the metabolic process is studied.
**Keywords:** mathematical model, metabolic process, self-organization, phase portrait, deterministic chaos, regular attractor, strange attractor, bifurcation, Poincaré section, Poincaré map, Lyapunov indices.

## 1. Introduction
In the present work, we continue the study of the mathematical model of the metabolic process in a cell *Arthrobacter globiformis*. It is based on the process of transformation of steroids in a bioreactor, which is well investigated in experiments [1]. The constructed mathematical model allows us to determine the internal and external parameters, with which the model describes the stationary modes of a bioreactor. The studies within the model showed that autooscillations must appear in the biochemical reaction under certain conditions [2-17]. These autooscillations predicted as early as in 1985 [2] were found experimentally in [18, 19].
Analogous autooscillations are observed in the processes of photosynthesis, glycolysis, variations of the calcium concentration in a cell, oscillations in heart muscle, and other biochemical systems [20-24].
The study of such autooscillations will allow one to investigate the internal dynamics of metabolic processes in cells, to find the structural-functional connections in a cell, by which its vital activity runs, and to clarify the evolution of the formation of these connections. The application of the mathematical apparatus of nonlinear dynamics to the study of metabolic processes will allow one to develop the general methods of synergetics considering the physical laws of self-organization in the Nature.

## 2. Mathematical Model
The mathematical model of the metabolic process running in a cell *Arthrobacter globiformis* at the transformation of steroids is constructed according to the general scheme of this process presented in Fig. 1. The model is based on the results of experimental studies of the process under flowing-through conditions with a fermenter in porous granules with immobilized cells *Arthrobacter globiformis* [3, 4].

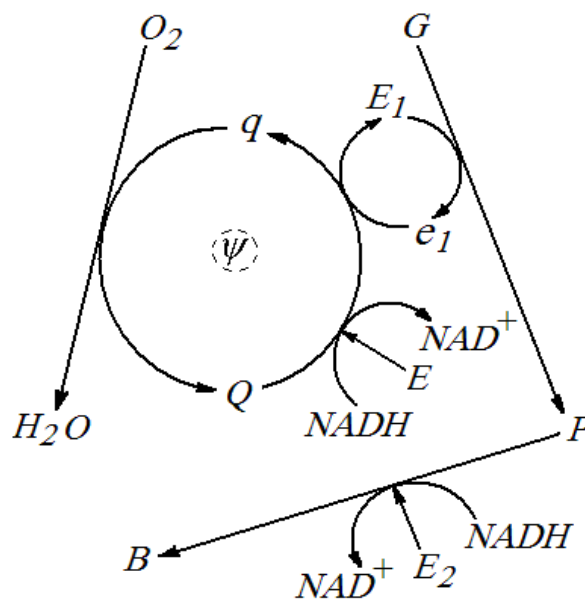

Fig. 1. General scheme of the metabolic process in a cell *Arthrobacter globiformis*.

*The variation of the concentration of hydrocortisone ($G$) is described by the equation*

$$\frac{dG}{dt} = \frac{G_0}{N_3 + G + \gamma_2 \psi} - l_1 V(E_1) V(G) - \alpha_3 G. \tag{1}$$

Under the action of the diffusion and the flow into pores of a macroporous granule to cells, hydrocortisone comes to the region of localization of the enzyme 3-ketosteroid-$\Delta$-dehydrogenase ($E_1$) (term $\frac{G_0}{N_3 + G + \gamma_2 \psi}$) and is transformed by this enzyme into prednisolone (term $l_1 V(E_1) V(G)$). A part of hydrocortisone is taken out from the biosystem by the flow (term $\alpha_3 G$). Here and below, the function $V(X)$ characterizes the adsorption of the enzyme in the region of local binding into active complexes; $V(X) = X/(1+X)$.

*The variation of the concentration of prednisolone ($P$):*

$$\frac{dP}{dt} = l_1 V(E_1) V(G) - l_2 V(E_2) V(N) V(P) - \alpha_4 P. \tag{2}$$

Prednisolone formed in the process (term $l_1 V(E_1) V(G)$) is transformed by the enzyme $20\beta$-oxysteroid-dehydrogenase ($E_2$) to its $20\beta$-oxyderivative (term $l_2 V(E_2) V(N) V(P)$). Under the action of a flow (term $\alpha_4 P$), a part of prednisolone goes out into the external solution.

*The variation of the concentration of $20\beta$-oxyderivative of prednisolone ($B$):*

$$\frac{dB}{dt} = l_2 V(E_2) V(N) V(P) - k_1 V(\psi) V(B) - \alpha_5 B. \tag{3}$$

The increase of the concentration of $B$ occurs as a result of the transformation of prednisolone (term $l_2 V(E_2) V(N) V(P)$). Its decrease is due to the use of $20\beta$-oxyderivative by cells in one of the possible modifications of the Krebs cycle (term $k_1 V(\psi) V(B)$), which increases the level of $NAD \cdot H$. Under the action of a flow (term $\alpha_5 B$), $B$ is washed out into the external solution.

*The variation of the concentration of the oxidized form of 3-ketosteroid-$\Delta$-dehydrogenase ($E_1$):*

$$\frac{dE_1}{dt} = E_{10} \frac{G^2}{\beta_1 + G^2} (1 - \frac{P + mN}{N_1 + P + mN}) -$$

$$-l_1 V(E_1)V(G) + l_4 V(e_1)V(Q) - \alpha_1 E_1. \qquad (4)$$

The biosynthesis of the enzyme is described by the term $E_{10} \dfrac{G^2}{\beta_1 + G^2}(1 - \dfrac{P + mN}{N_1 + P + mN})$, which is defined by the activation by the substrate $G$ and the inhibition by the reaction products $P$ and $N$. The decrease of the concentration of this form of the enzyme in the process of transformation of hydrocortisone is given by the term $l_1 V(E_1)V(G)$, and its increase in the process of reduction of the respiratory chain corresponds to the term $l_4 V(e_1)V(Q)$. The inactivation of the enzyme due to the proteolysis is described by the term $\alpha_1 E_1$.

*The variation of the concentration of the reduced form of 3-ketosteroid-$\Delta$-dehydrogenase ($e_1$):*

$$\frac{de_1}{dt} = -l_4 V(e_1)V(Q) + l_1 V(E_1)V(G) - \alpha_1 e_1. \qquad (5)$$

Its level decreases in the process of reduction of the respiratory chain (term $-l_4 V(e_1)V(Q)$) and due to the inactivation (term $\alpha_1 e_1$) and increases at the transformation of hydrocortisone (term $l_1 V(E_1)V(G)$).

*The variation of the level of the oxidized form of the respiratory chain ($Q$)*

$$\frac{dQ}{dt} = 6lV(2-Q)V(O_2)V^{(1)}(\psi) - l_6 V(e_1)V(Q)_1 - l_7 V(Q)V(N), \qquad (6)$$

where $V^{(1)}(\psi) = 1/(1 + \psi^2)$. We accept that the concentration of menaquinone $Q^0 + q^0 = 2$, where $q$ is the reduced form of the respiratory chain.

The respiratory chain is oxidized by oxygen (term $6lV(2-Q)V(O_2)V^{(1)}(\psi)$) and is reduced with the help of $e_1$ (term $-l_6 V(e_1)V(Q)$) and due to the high level of $NAD \cdot H$ (term $-l_7 V(Q)V(N)$).

*The variation of the concentration of oxygen ($O_2$):*

$$\frac{dO_2}{dt} = \frac{O_{20}}{N_5 + O_2} - lV(2-Q)V(O_2)V^{(1)}(\psi) - \alpha_7 O_2. \qquad (7)$$

Under the action of a flow (terms $\dfrac{O_{20}}{N_5 + O_2}$ and $\alpha_7 O_2$), the level of aeration of a cell is changed. The concentration of oxygen decreases at the oxidation of the respiratory chain (term $-lV(2-Q)V(O_2)V^{(1)}(\psi)$).

*The variation of the concentration of $20\beta$-oxysteroid-dehydrogenase ($E_2$):*

$$\frac{dE_2}{dt} = E_{20}\frac{P^2}{\beta_2 + P^2}\frac{N}{\beta + N}(1 - \frac{B}{N_2 + B}) -$$

$$- l_{10}V(E_2)V(N)V(P) - \alpha_2 E_2 \qquad (8)$$

The increase of the level of the given enzyme occurs due to the biosynthesis: $E_{20}\dfrac{P^2}{\beta_2 + P^2}\dfrac{N}{\beta + N}(1 - \dfrac{B}{N_2 + B})$. Prednisolone and $NAD \cdot H$ are activators of this process, and $20\beta$-oxyderivative is an inhibitor. The decrease of the level of the given enzyme occurs as a result of the inactivation ($-\alpha_2 E_2$) and the process of transformation of prednisolone ($-l_{10}V(E_2)V(N)V(P)$).

*The variation of the concentration of $NAD \cdot H$ ($N$) of a cell:*

$$\frac{dN}{dt} = -l_2 V(E_2)V(N)V(P) - l_7 V(Q)V(N) +$$

$$+ k_2 V(B)\frac{\psi}{K_{10} + \psi} + \frac{N_0}{N_4 + N} - \alpha_6 N. \qquad (9)$$

The level of the co-enzyme $N$ decreases in the process of transformation $P \Rightarrow B$, in the process of reduction of the respiratory chain ($-l_7 V(Q)V(N)$), and due to a flow ($-\alpha_6 N$). It increases at the use of $B$ by cells in the Krebs cycle as a substrate ($k_2 V(B) \frac{\psi}{K_{10}+\psi}$) and in the presence of endogenous substrates ($\frac{N_0}{N_4+N}$) in the environment.

*The variation of the level of kinetic membrane potential ($\psi$):*

$$\frac{d\psi}{dt} = l_5 V(E_1)V(G) + l_8 V(N)V(Q) - \alpha\psi . \tag{10}$$

The kinetic membrane potential arises at the transformation of hydrocortisone ($l_5 V(E_1)V(G)$) and the reduction of the respiratory chain ($l_8 V(N)V(Q)$) at a high level of $NAD \cdot H$ and decreases due to other metabolic processes ($-\alpha\psi$). The variation of the level of $\psi$ changes its regulatory role (1), (3), (6), (7), (9). If the potential is high, the respiratory chain is blocked and held in the reduced state.

The main parameters of the system, with which we fit the relevant experimental data, are as follows: $l = l_1 = k_1 = 0.2$; $l_2 = l_{10} = 0.27$; $l_5 = 0.6$; $l_4 = l_6 = 0.5$; $l_7 = 1.2$; $l_8 = 2.4$; $k_2 = 1.5$; $E_{10} = 3$; $\beta_1 = 2$; $N_1 = 0.03$; $m = 2.5$; $\alpha = 0.033$; $a_1 = 0.007$; $\alpha_1 = 0.0068$; $E_{20} = 1.2$; $\beta = 0.01$; $\beta_2 = 1$; $N_2 = 0.03$; $\alpha_2 = 0.02$; $G_0 = 0.019$; $N_3 = 2$; $\gamma_2 = 0.2$; $\alpha_5 = 0.014$; $\alpha_3 = \alpha_4 = \alpha_6 = \alpha_7 = 0.001$; $O_{20} = 0.015$; $N_5 = 0.1$; $N_0 = 0.003$; $N_4 = 1$; $K_{10} = 0.7$.

The study of solutions of the given mathematical model was carried out with the help of the theory of nonlinear differential equations [25-27].

In the numerical solution of this autonomous system of nonlinear differential equations, we used the Runge--Kutta--Merson method. The accuracy of calculations was set to be $10^{-8}$. To attain the reliability of a solution, when the system passes from the initial transient phase onto the asymptotic solution with an attractor, the duration of calculations was taken to be $10^6$. For this time interval, the trajectory "sticks" onto the appropriate attractor.

The various types of autooscillatory modes are studied with the help of the construction of exact phase-parametric diagrams. We found the scenarios of appearance of bifurcations at the transition of the dynamical process from one type of an attractor to another one. For the most characteristic modes, we calculated the total spectra of Lyapunov indices (Table 1).

To construct a phase-parametric diagram, we used the method of section. In the phase space of trajectories of the system, we place a cutting plane with $P = 0.2$. Such choice is explained by the symmetry of oscillations relative to this point of this variable in multiple modes. If the trajectory $P(t)$ crosses this plane in a certain direction, we mark the value of chosen variable (e.g., $G$) on the phase-parametric diagram. In such way, we have the point corresponding to the section of a trajectory by the two-dimensional plane. If the multiple periodic limiting cycle appears, we obtain a number of points, which will be coincide in a period. If a deterministic chaos arises, the points of intersection of trajectories by the plane will be placed chaotically.

In order to uniquely identify the form of an attractor for the chosen points, we calculated the total spectrum of Lyapunov indices and determined their sum $\Lambda = \sum_{j}^{10} \lambda_j$ (see Table 1). The calculation was carried out by Benettin's algorithm with orthogonalization of the vectors of perturbation by the Gram--Schmidt method [26, 28, 29].

### 3. Results of Studies

We now consider the dynamics of modes within the mathematical model (1)-(10) under a variation of the dissipation of a kinetic membrane potential $\alpha$ (10) [16, 17]. We found the autooscillatory and chaotic modes with various multiplicities. The projections of their phase portraits have a characteristic form shown in Fig. 2,a,b.

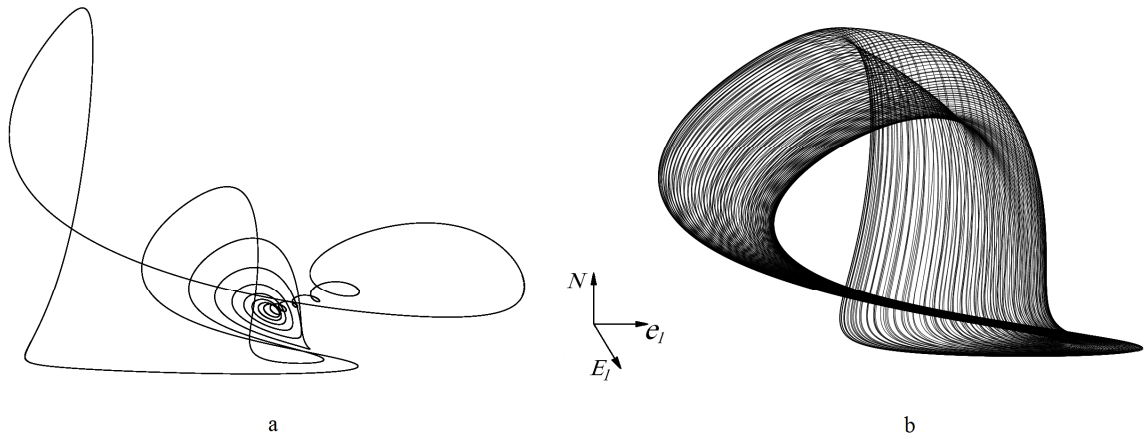

Fig. 2. Projections of the phase portraits of regular attractors;

a – autoperiodic cycle $14 \cdot 2^0$ for $\alpha = 0.033$;

b – quasiperiodic cycle $\approx 31 \cdot 2^0$ for $\alpha = 0.0321375$.

Let us consider a part of the bifurcation diagram not studied earlier. In Fig. 3, we show the bifurcation diagram for $\alpha \in$ (0.032159, 0.032166).

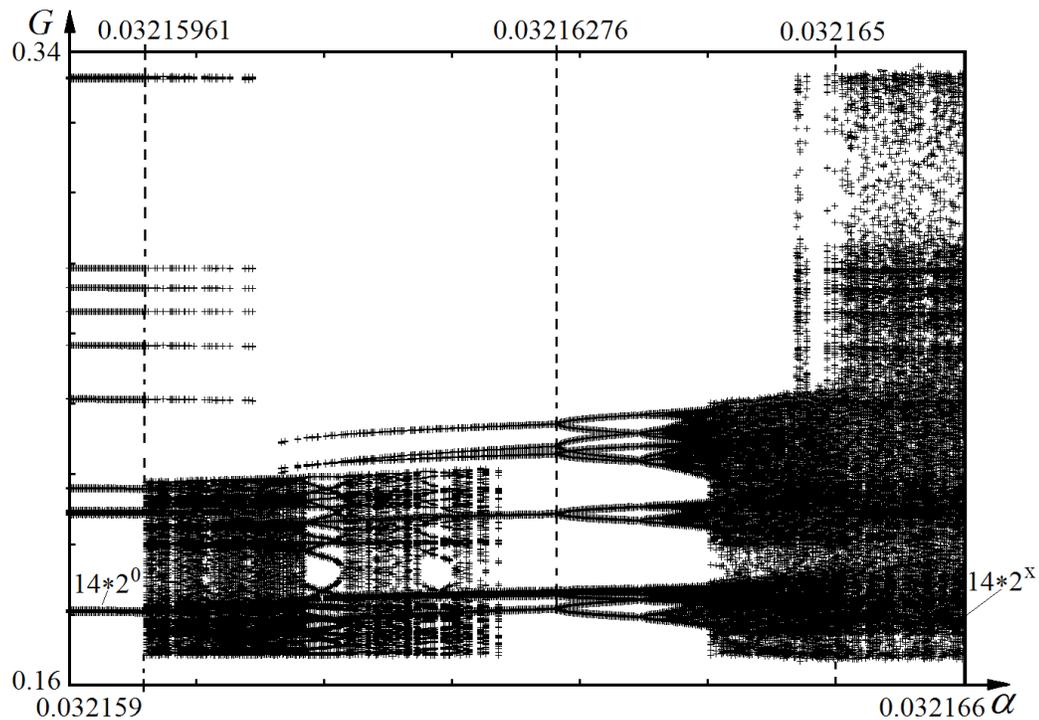

Fig. 3. Bifurcation diagram of the system for $\alpha \in$ (0.032159, 0.32166).

For $\alpha \in$ (0.0321590, 0.03215960), the regular attractor of the 14-fold period $14 \cdot 2^0$ is kept in the system. For $\alpha = 0.03215961$, we observe the appearance of the period doubling bifurcation with the generation of the regular attractor $14 \cdot 2^1$ (Table 1). Then for $\alpha = 0.03215962$, there arises the bifurcation of the generation of a two-dimensional torus (the Neimark bifurcation). The configuration of kinetic curves is instantly changed, and the quasiperiodic attractor with $n$-fold period is established on the toroidal surface $\approx n \cdot 2^0 (t)$ (Figs. 4,a and 5,a).

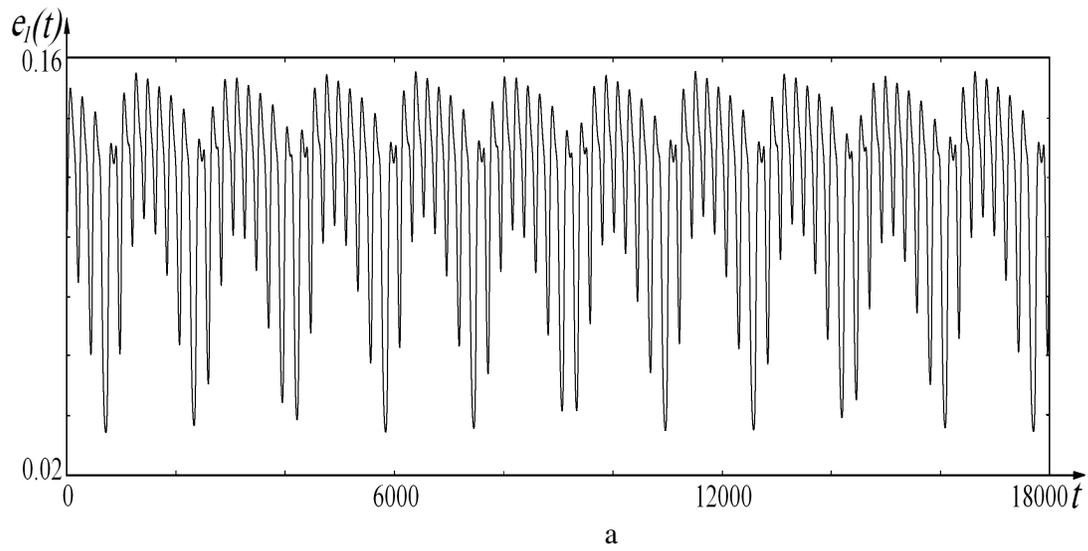

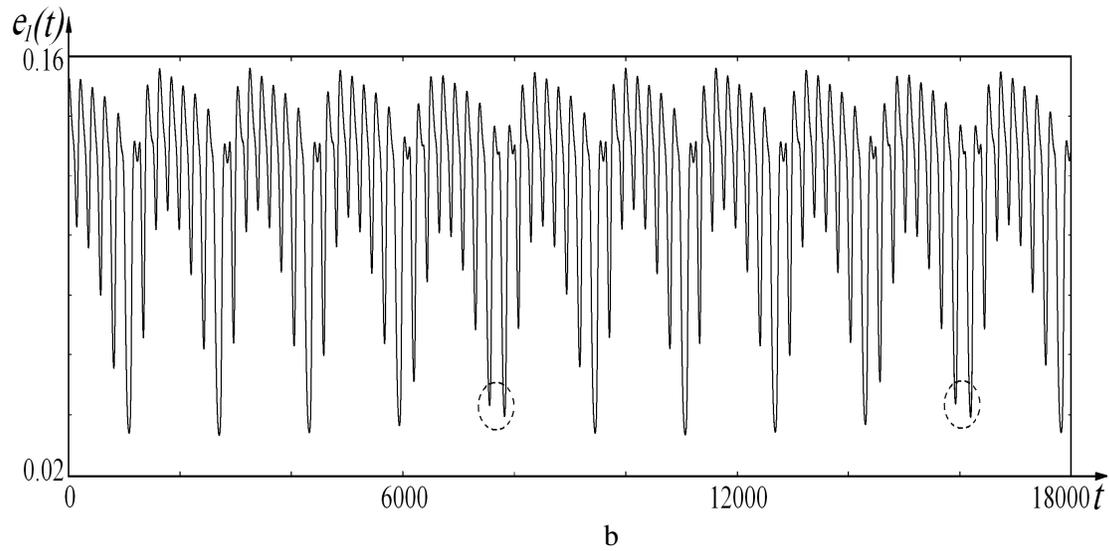

Fig. 4. Kinetic curve of the variable $e_1(t)$;

a - regular attractor of the quasiperiodic cycle $\approx n*2^0$ on the toroidal surface for $\alpha = 0.03215962$.

b - regular attractor $36 \cdot 2^0$ for $\alpha = 0.032162$.

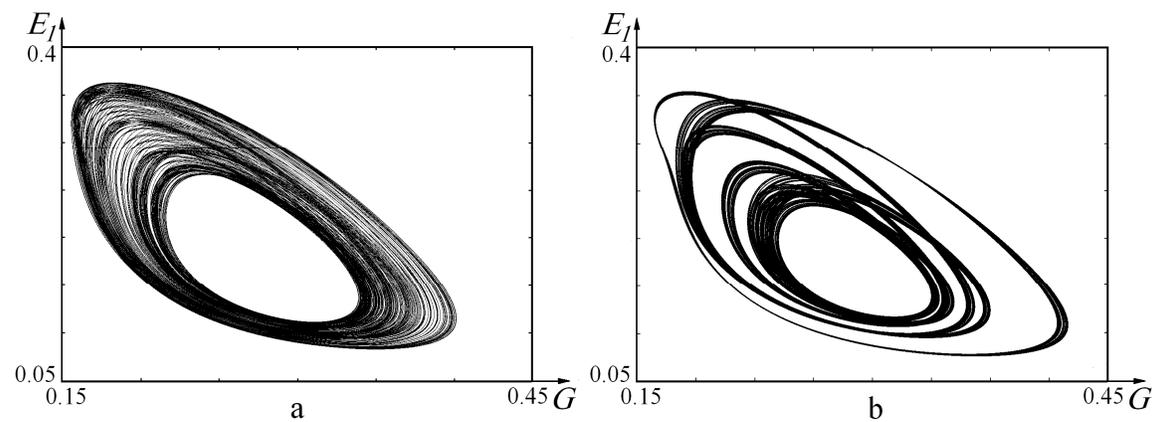

Fig. 5. Projections of phase portraits;

a – regular attractor of the quasiperiodic cycle $\approx n \cdot 2^0$ on the toroidal surface for $\alpha = 0.03215962$;

b – strange attractor $7 \cdot 2^x$ for $\alpha = 0.032164$.

As $\alpha$ increases, the given attractor loses the stability, by passing periodically to the 14-fold limiting cycle ($\alpha = 0.032160$), which corresponds to the gaps in Fig. 3,a. In addition, other various multiple modes arise. For example, for $\alpha = 0.032161$, $0.0321615$, and $0.032162$, the regular attractors $29 \cdot 2^0$, $7 \cdot 2^0$, and $36 \cdot 2^0$ appear, respectively (Fig. 4,b). As $\alpha$ increases, we see the appearance of bifurcations of the limiting cycle. Moreover, the instant structural rearrangement of the type "order-order" occurs; i.e., as a result of the self-organization, the regular attractor of some form is replaced instantly by a regular attractor of some other form. In this case, the trajectories leave the region of attraction of the attractor and are drawn in the region of attraction of another regular attractor.

The interesting scenario of the metabolic process is observed in the interval $\alpha \in (0.0321626, 0.032164)$. In Fig. 6, we present a magnified part of the bifurcation diagram in Fig. 3.

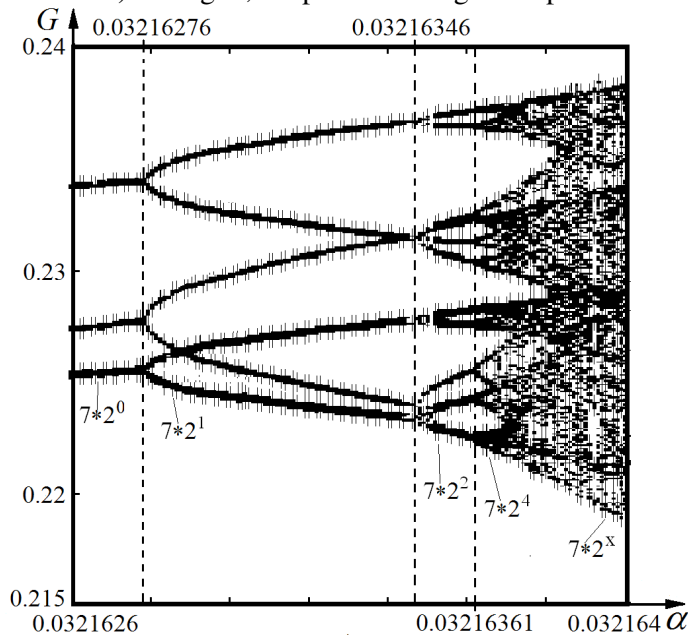

Fig. 6. Phase-parametric diagram of the system for $\alpha \in (0.0321626, 0.32164)$, where Feigenbaum's scenario is observed.

At the beginning of the interval at $\alpha = 0.0321626$, the regular attractor $7 \cdot 2^0$ is formed on the toroidal surface. For $\alpha_j = 0.03216276$, the bifurcation yields the doubling of the period, and the regular attractor $7 \cdot 2^1$ arises on the toroidal surface. For $\alpha_{j+1} = 0.03216346$ and $\alpha_{j+2} = 0.03216361$, we see the attractors $7 \cdot 2^2$ and $7 \cdot 2^4$, respectively. This sequence of bifurcations satisfies the relation

$$\lim_{t \to \infty} \frac{\alpha_{j+1} - \alpha_j}{\alpha_{j+2} - \alpha_{j+1}} \approx 4.667.$$

This number is very close to Feigenbaum's universal constant $\delta = 4.669211660910...$ characterizing the infinite cascade of bifurcations at the transition to a deterministic chaos. Thus, as the coefficient of dissipation $\alpha$ increases in this region, the period of a complicated regular attractor on the torus is doubled by Feigenbaum's scenario [37-40].

The further increase in $\alpha$ causes a deviation from the given scenario and the formation of the strange attractor $7 \cdot 2^x$ ($\alpha = 0.032164$, Fig. 5,b) as a result of the intermittency. But then, for $\alpha = 0.032174$, the strange attractor $14 \cdot 2^x$ appears (Fig. 7,b). In the interval $\alpha = (0.032164, 0.032174)$ as a result of the intermittency of these chaotic cycles, we observe the transition between them: $(7 \leftrightarrow 14) \cdot 2^x$. In Fig. 7,a for $\alpha = 0.032165$, we show a projection of the phase portrait of a mutual transition of the given strange attractors. Figure 8 presents the kinetic curve for the variable

$e_1(t)$ for tis mode. We observe the transition "chaos-chaos": $(7 \leftrightarrow 14) \cdot 2^x$. Moreover, the strange attractor $7 \cdot 2^x$ on the left and the strange attractor $14 \cdot 2^x$ on the right move toward each other. Since there are no other attractors of the system in this region, the trajectory is chaotically kept in the region of attraction of the strange attractor $14 \cdot 2^x$ or the strange attractor $7 \cdot 2^x$ Under the effect of bifurcations, the trajectory is aperiodically drawn in one of the regions of the given strange attractors after the transient process. According to the values of higher Lyapunov indices (Table 1), the formed limiting set is unstable by Lyapunov.

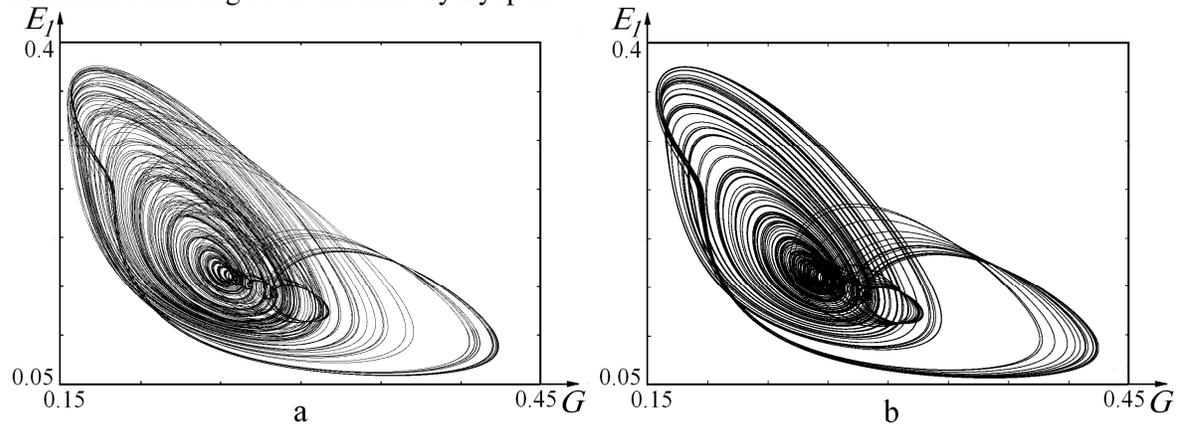

Fig. 7. Projections of the phase portraits;
a – strange attractor of the mutual transition $(7 \leftrightarrow 14) \cdot 2^x$ for $\alpha = 0.032165$;
b – strange attractor $14 * 2^x$ for $\alpha = 0.032174$.

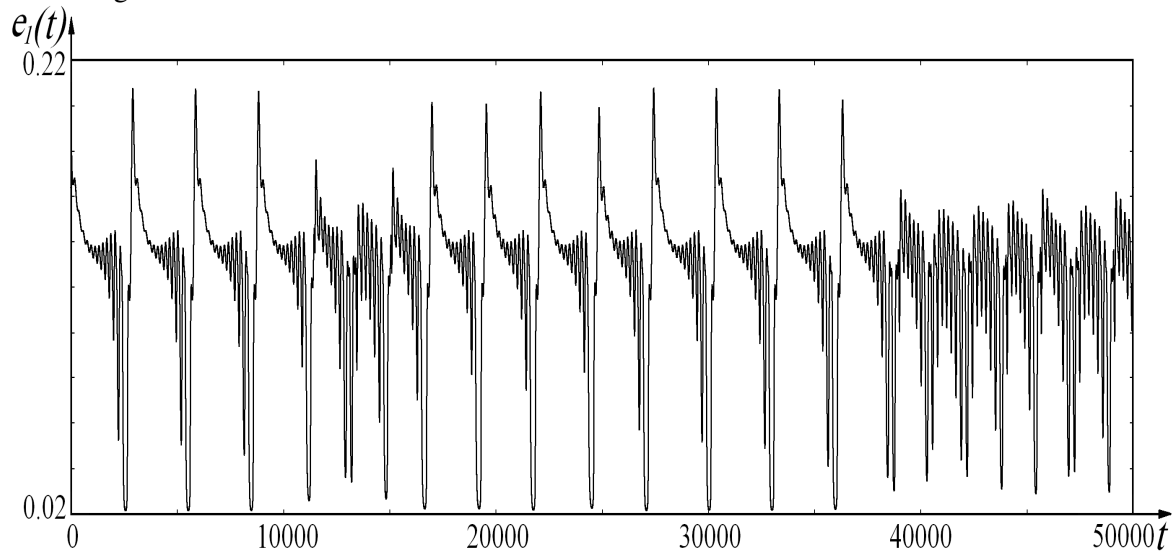

Fig. 8. Kinetic curve of the variable $e_1(t)$ of the mutual transition of the strange attractors $(7 \leftrightarrow 14) \cdot 2^x$ for $\alpha = 0.032165$.

For the given strange attractor, we constructed a projection of the section by the plane $P = 0.2$ and the Poincaré map in Fig. 9,a,b. The choice of a cutting surface was made to attain the maximum number of intersections of the given component and the phase trajectory $P(t)$, as the former decreases, without contacts.

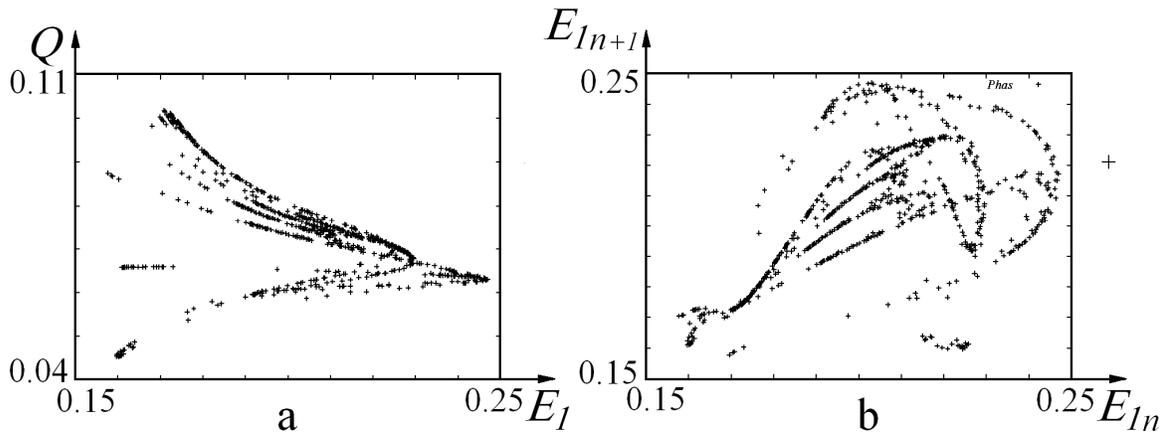

Fig. 9. Projection of the section by the plane $P = 0.2$ (a) and Poincaré map (b) of the strange attractor formed during the mutual transition $(7 \leftrightarrow 14) \cdot 2^x$ for $\alpha = 0.032165$.

The obtained points of intersections and the Poincaré maps are grouped along several curves that form a geometric self-similarity. On the projection, we see clearly the fractality of this strange attractor. In addition, these curves do not create a quasistrip structure. Their number increases permanently with the duration of numerical integration of the system. This testifies to the impossibility of any reduction of the given complicated mathematical model to some one-dimensional discrete approximation without loss of the information about the dynamics of the metabolic process in a cell. We note that the general scheme (Fig. 2) includes only the main parts of the metabolic process running in any cell with substrate-enzyme reactions and in the respiratory chain. Therefore, the model gives a rather general qualitative representation of the dynamics of the internal self-organization of the metabolic process in a cell.

Table 1. Total spectra of Lyapunov indices for attractors of the system under study ($\lambda_4 - \lambda_{10}$ are not important for our investigation).

| $\alpha$ | Attractor | $\lambda_1$ | $\lambda_2$ | $\lambda_3$ | $\Lambda$ |
|---|---|---|---|---|---|
| 0.0321590 | $14 \cdot 2^0$ | .000056 | -.000214 | -.003250 | -.898509 |
| 0.0321596 | $14 \cdot 2^0$ | .000040 | -.000142 | -.003306 | -.898550 |
| 0.03215961 | $14 \cdot 2^1$ | .000078 | -.000150 | -.003394 | -.899865 |
| 0.03215962 | $\approx n \cdot 2^0 (t)$ | .000063 | .000026 | -.000274 | -.905553 |
| 0.032160 | $14 \cdot 2^0$ | .000040 | -.000146 | -.003365 | -.899368 |
| 0.032161 | $29 \cdot 2^0$ | .000051 | -.000142 | -.000123 | -.905352 |
| 0.0321615 | $7 \cdot 2^0$ | .000062 | -.000596 | -.000576 | -.902277 |
| 0.032162 | $36 \cdot 2^0$ | .000064 | -.000171 | -.000155 | -.905320 |
| 0.0321626 | $7 \cdot 2^0 (t)$ | .000063 | -.000097 | -.001180 | -.902078 |
| 0.03216276 | $7 \cdot 2^1 (t)$ | .000062 | -.000005 | -.001267 | -.902189 |
| 0.03216346 | $7 \cdot 2^2 (t)$ | .000047 | .000025 | -.001252 | -.902056 |
| 0.03216361 | $7 \cdot 2^3 (t)$ | .000048 | -.000023 | -.001265 | -.902267 |
| 0.032164 | $7 \cdot 2^x$ | .000367 | .000018 | -.001641 | -.902164 |
| 0.032165 | $(7 \leftrightarrow 14) \cdot 2^x$ | .000363 | -.000004 | -.001598 | -.904005 |

| 0.032174 | $14 \cdot 2^x$ | .000693 | .000020 | -.003534 | -.901422 |

**4. Conclusions**
We have constructed a mathematical model of the metabolic process in a cell *Arthrobacter globiformis* at the transformation of steroids. With the help of the given model, we have found the autooscillations in agreement with experiment, which show the complicated internal dynamics in a cell. The model is optimized by the number of variables of the system required for a qualitative description of the metabolic process under study. The given model involves the general regularities characteristic of any cell consuming a substrate, on the whole. The autooscillations arise on the level of the substrate-enzyme interaction with participation of the redox process in the respiratory chain and characterize the times of such interactions. At the synchronization of the given processes, the autooscillations characterizing the self-organization of the metabolic process on the whole are observed. At the desynchronization of the given processes, we see the adaptation of the metabolic process in a cell to varying external conditions in the environment with conservation of its functionality. The scenario of the transitions "order-chaos", "chaos-order", "order-order", and "chaos-chaos" is studied with the help of Poincaré sections and maps. The total spectra of Lyapunov indices are calculated, and the structural stability of the obtained attractors is studied. Feigenbaum's scenario and the Neimark bifurcation are found. The results will allow one to carry on the search for metabolic oscillations in a cell and to clarify the physical laws of self-organization.

**Acknowledgement**
The work is supported by the project N 0112U000056 of the National Academy of Sciences of Ukraine.